# Shapes of rotating normal fluid $^3$He versus superfluid $^4$He droplets in molecular beams

Deepak Verma[1], Sean M. O'Connell[1], Alexandra J. Feinberg[1], Swetha Erukala[1], Rico M. Tanyag[1,2], Charles Bernando[3,4], Weiwu Pang[4], Catherine A. Saladrigas[6,7], Benjamin W. Toulson[6], Mario Borgwardt[6], Niranjan Shivaram[8,9], Ming-Fu Lin[8], Andre Al Haddad[10], Wolfgang Jäger[11], Christoph Bostedt[10,12], Peter Walter[8], Oliver Gessner[6*] and Andrey F. Vilesov[1,3*]

[1] Department of Chemistry, University of Southern California, Los Angeles, California 90089, USA
[2] Institute for Optics and Atomic Physics, Technical University of Berlin, Berlin 10623, Germany
[3] Department of Physics and Astronomy, University of Southern California, Los Angeles, California 90089, USA
[4] School of Information Systems, BINUS University, Jl. K.H. Syahdan No. 9, Kemanggisan, Palmerah, Jakarta 11480 Indonesia
[5] Viterbi School of Engineering, University of Southern California, Los Angeles, California 90089, USA
[6] Chemical Sciences Division, Lawrence Berkeley National Laboratory, Berkeley, California 94720, USA
[7] Department of Chemistry, University of California, Berkeley, California 94720, USA
[8] LCLS, SLAC National Accelerator Laboratory, 2575 Sand Hill Road, Menlo Park, California 94025, USA
[9] Department of Physics and Astronomy, Purdue University, West Lafayette, Indiana 47907, USA
[10] Laboratory for Femtochemistry (LSF), Paul Scherrer Institut, 5232 Villigen-PSI, Switzerland
[11] Department of Chemistry, University of Alberta, Edmonton, Alberta T6G 2G2, Canada
[12] LUXS Laboratory for Ultrafast X-ray Sciences, Institute of Chemical Sciences and Engineering, École Polytechnique Fédérale de Lausanne (EPFL), CH-1015, Lausanne, Switzerland

Present Address:

[⊥] OVO (PT. Visionet Internasional), Lippo Kuningan 20$^{th}$ floor, Jl. HR Rasuna Said Kav. B-12, Setiabudi, Jakarta, 12940, Indonesia
[#] Department of Physics and Astronomy, Purdue University, West Lafayette, Indiana 47907, USA

March 9$^{th}$, 2020

**Abstract**

Previous single-pulse extreme ultraviolet and X-ray coherent diffraction studies revealed that superfluid $^4$He droplets obtained in free jet expansion acquire sizable angular momentum, resulting in significant centrifugal distortion. Similar experiments with normal fluid $^3$He droplets may help elucidating the origin of the of the large degree of rotational excitation and highlight similarities and differences of dynamics in normal and superfluid droplets. Here, we present the first comparison of the shapes of isolated $^3$He and $^4$He droplets following expansion of the corresponding fluids in vacuum at temperatures as low as ~ 2 K. Large $^3$He and $^4$He droplets with average radii of ~160 nm and ~350 nm, respectively, were produced. We find that the majority of the $^3$He droplets in the beam correspond to rotating oblate spheroids with reduced average angular momentum ($\Lambda$) and reduced angular velocities ($\Omega$) similar to that of $^4$He droplets. Given the different physical nature of $^3$He and $^4$He, this similarity in $\Lambda$ and $\Omega$ may be surprising and suggest that similar mechanisms induce rotation regardless of the isotope. We hypothesized that the observed distribution of droplet sizes and angular momenta stem from processes in the dense region close to the nozzle. In this region, the significant velocity spread and collisions between the droplets induce excessive rotation followed by droplet fission. The process may repeat itself several times before the droplets enter the collision-fee high vacuum region further downstream.

# 1. Introduction

Bosonic superfluid helium-4 ($^4$He) droplets, produced in molecular beams, constitute a versatile medium for experiments in physics and chemistry. Notably, droplets consisting of a few thousand He atoms are frequently used as ultra-cold matrices for the spectroscopic interrogation of single molecules, radicals, ionic species, and diverse clusters [1-10]. Single molecules embedded in $^4$He droplets can also provide a unique probe for superfluidity on atomic-length scales via renormalization of molecular rotational constants [11-15]. More recently, experiments with superfluid $^4$He have been extended to much larger droplets, containing up to ~$10^{11}$ atoms, and ranging in diameter from hundreds of nanometers up to a few micrometers [16, 17]. Single droplets in this size range have been studied by ultrafast coherent scattering using femtosecond X-ray and XUV pulses from free electron lasers (FEL) and intense, laboratory-based high-order harmonics sources [18-23]. It was found that large $^4$He droplets have sizable angular momentum and are subject to considerable centrifugal distortion [18, 21-23]. Rotation of superfluid $^4$He droplets is associated with the creation of quantum vortices, a physical manifestation of quantized angular momentum in these bosonic species [24-27]. Quantum vortices inside $^4$He droplets have been visualized by doping them with a large number of xenon (Xe) atoms. The dopants are attracted by the vortices, leading to aggregation around the vortex cores and the formation of filament-shaped clusters [18-20].

Experiments involving droplets of the rare fermionic helium-3 isotope ($^3$He) have also been performed [11, 28-36]. While $^3$He may exist as a superfluid at temperatures T $\lesssim$ 1 mK, it constitutes of a normal fluid under typical molecular beam temperatures of ≈ 0.15 K [37, 38]. Recent density functional calculations show that the rotating $^3$He should follow corresponding classical shapes [39]. It is important to expand X-ray imaging experiments to rotating $^3$He droplets to enable a direct comparison of droplet shapes and rotational properties for the two quantum fluids. The comparison of the angular momenta and angular velocities of droplets consisting of two different isotopes may also shed light on the origin of rotation in droplets produced via fluid expansion into vacuum, which remains obscure.

In molecular beam experiments, He droplets are produced by expanding pressurized He through a cryogenic nozzle into vacuum [2, 8, 9, 16]. Figure 1(a) and (b) illustrate the production of $^4$He and $^3$He droplets via corresponding pressure-temperature (P-T) phase diagrams. The adiabatic expansion proceeds along isentropes, i.e., lines of constant entropy, starting at an initial condition defined by the nozzle temperature, $T_0$, and a stagnation pressure, $P_0$ and ending at a set of final values $T_f$, $P_f$ on the saturated vapor pressure (SVP) curve. Large droplets are obtained when the isentrope crosses the SVP curve from the fluid side, corresponding to boiling of the fluid and its fragmentation into droplets. Larger droplets are obtained at lower $T_f$, at which less violent boiling leaves larger droplets intact. In vacuum, the temperature of the droplets further decreases via evaporative cooling down to 0.15 K and 0.38 K for $^3$He [37, 38] and $^4$He, [11, 40] respectively. It is worth noticing that $^4$He becomes superfluid below 2.17 K. In the superfluid state, boiling ceases, which may lend further stability to the insipient $^4$He droplets. Due to fast evaporative cooling [8], $^4$He droplets become superfluid close to the nozzle, however the location of the superfluid transition as well as the kinetics of the droplet cooling upon expansion remain unknown.

In this article, we report on the characterization of $^3$He droplets produced with nozzle temperatures as low as $\approx$ 2 K. Using ultrafast X-ray scattering at an X-FEL, the properties of individual, free $^3$He and $^4$He droplets are analyzed and compared, in particular, with respect to their size, shape and angular momenta. A wide range of $^3$He and $^4$He droplets sizes are obtained with average radii of 162 nm and 355 nm, respectively. The aspect ratio of droplets from both isotopes are found to have similar average values of 1.055 for $^3$He and 1.076 for $^4$He. Accordingly, the reduced angular momentum and reduced angular velocity in $^3$He and $^4$He droplets are comparable. Comparison of the results obtained with $^4$He and $^3$He at different expansion conditions may help to gain a better understanding of the mechanism underlying the production of rotating droplets in free nozzle beam expansion sources.

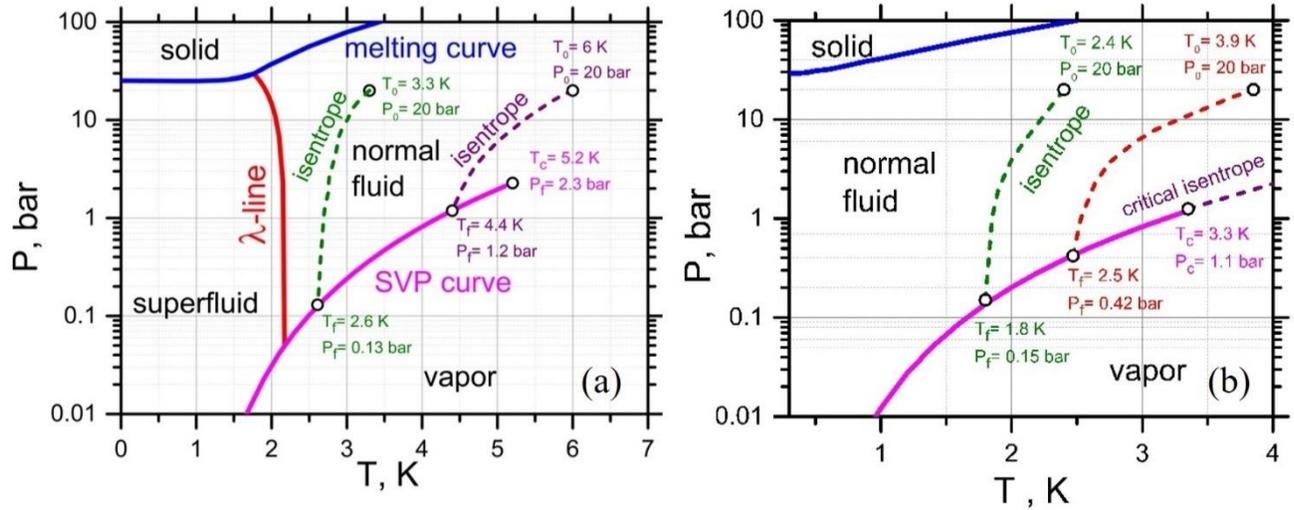

**Figure 1**. P-T phase diagrams for (a) $^4$He and (b) $^3$He. The diagrams are based on refs. [41, 42] for $^4$He and [43, 44] for $^3$He. The pink SVP curves mark the saturated vapor pressure boundaries.

## 2. Experimental

He droplets are produced by expanding pressurized He through a cryogenic nozzle with a 5 μm orifice into vacuum, the details of which are described elsewhere [2, 8, 9, 16]. Considering the lower critical point of $^3$He ($T_C$ = 3.3 K, $P_C$ = 1.1 atm) compared to that of $^4$He ($T_C$ = 5.2 K, $P_C$ = 2.3 atm), lower nozzle temperatures are required to obtain $^3$He droplets of the same sizes as those consisting of $^4$He. For example, for $P_0$ = 20 bar, $^4$He droplets with an average size of $\langle N_4 \rangle = 10^7$ are produced at $T_0 = 7$ K, [16] while $T_0 = 5$ K is required to obtain $^3$He droplets of the same size $\langle N_3 \rangle$ [31-33, 44]. The temperature difference of 2 K correlates well with the corresponding difference in critical temperatures of the two isotopes. In a spherical droplet, the radius and the number of the atoms it contains are related by $R_{3,4} = \beta_{3,4} \cdot \sqrt[3]{N_{3,4}}$, where the coefficient $\beta$ can be obtained from the number density of the corresponding liquid to be 0.245 nm and 0.222 nm in $^3$He and $^4$He droplets, respectively [38, 42]. Large $^4$He droplets can be produced with modern closed-cycle refrigerators that can reach temperatures down to ≈ 3.5 K. However, to reach the lower temperatures required to produce large $^3$He droplets, we instead employ a liquid helium flow cryostat with a cooling power of up to 1 W at 1.8 K. $^3$He and $^4$He droplets are produced at constant $P_0$ = 20 bar and varying $T_0$, ranging from 2 to 4.5 K.

Due to the considerable cost of $^3$He gas, a recycling system is employed during the experiments as described in the Supplementary Material (SM) [45]. Filling the gas handling system requires about 10 L·bar of room temperature $^3$He. For comparison, at standard operating conditions ($T_0$ = 3 K, $P_0$ = 20 bar), the flow rate of the He gas is ~3 cm$^3$·bar/s and the filling amount of gas would only be sufficient for about 1 hour of operation. During the experiments, $^3$He gas is continuously collected from the exhausts of the backing scroll pumps, purified in a liquid nitrogen cooled zeolite trap, pressurized by a metal membrane compressor and resupplied to the nozzle with minimal losses. The $^3$He gas used is 99.9% pure with the remaining 0.1% impurity being mostly $^4$He.

The experiments are performed using the LAMP end station at the Atomic, Molecular and Optical (AMO) instrument of the Linac Coherent Light Source (LCLS) XFEL. [46, 47] The focused XFEL beam (~2 μm full-width-at-half-maximum, FWHM) intersects the He droplet beam ~70 cm downstream from the nozzle. The XFEL is operated at 120 Hz, a photon energy of 1.5 keV ($\lambda$ = 0.826 nm), a pulse energy of ~1.5 mJ and a pulse duration of ~100 fs (FWHM). The small pulse length and large number of photons per pulse (~$10^{12}$) enable the capture of instantaneous shapes of individual droplets. Diffraction images are recorded with a pn-charge-coupled device (pnCCD) detector containing 1024×1024 pixels, each 75×75 μm$^2$ in size, which is centered along the XFEL beam axis ~735 mm downstream from the interaction point. The detector consists of two separate panels (1024×512 pixels each), located closely above and below the X-ray beam. Both panels also have a central, rectangular section cut-out to accommodate the primary X-ray beam. The diffraction patterns are recorded at small scattering angles and, thus, predominantly contain information on the column density of the droplets in the direction perpendicular to the detector plane.

## 3. Results

Figure 2 shows several diffraction patterns from pure $^3$He droplets. The images are characterized by sets of concentric contours. Images in Figures 2(a) and (b) exhibit a series of circular and elliptical contours, respectively, with different spacings. Figure 2(c), however, shows an elongated diffraction contour with pronounced streaks radiating away from the center. The collected diffraction patterns are characteristic of spherical (Figure 2(a)), and spheroidal (oblate) or capsule (prolate) (Figure 2(b) and(c)) droplet shapes, as previously observed in $^4$He droplets [18, 23, 48]. Spheroidal and prolate shapes, in particular, result from the centrifugal deformation of droplets with considerable angular momentum.

The droplet shapes are characterized by the distances between the center and the surface in three mutually perpendicular directions: $a \geq b \geq c$. For an oblate axisymmetric droplet, $a = b > c$, with $c$ along the rotation axis, whereas $a > b > c$ in the case of triaxial prolate shapes with $c$ along the rotation axis [16, 21]. The observed diffraction patterns do not provide direct access to the actual values of $a$, $b$ and $c$, due to the droplets' unknown orientations with respect to the X-ray beam. Instead, the images are characterized by two semi-axes of the projection of a droplet onto the detector plane, which will be referred to as $A$ and $C$ ($A > C$), corresponding to a projection aspect ratio, $AR = A / C$. For an axisymmetric droplet with an unknown orientation with respect to the X-ray beam, the value of A corresponds to the $a$-axis, whereas the value of $C$ only constitutes an upper bound for the $c$-axis. In the case of a triaxial droplet, the value of $A$ gives a lower bound for the $a$-axis, whereas the value of $C$ gives a lower bound for the $b$-axis and an upper bound for the $c$-axis. In this section, we will discuss the experimental results in terms of the apparent $A$, $C$ and $AR$ values, from which the average actual sizes of the axisymmetric droplets are obtained. The values of A and C are obtained from the diffraction patterns as described elsewhere (supplementary material in Reference [21]).

The values of the half axes $A$ and $C$, as well as their $ARs$, are noted for each panel in Figure 2. The calculated $A$ and $C$ values from Figure 2(a) are very similar (within ~ 3%), indicative of a spherical droplet shape or a spheroid with its symmetry axis aligned perpendicular to the detector plane. The diffraction pattern shown in Figure 2(b) originates from a larger droplet (larger half-axis values). Here, the two half-axes differ by ~ 34% ($AR = 1.34$), which is indicative of a spheroidal or ellipsoidal droplet. The streaked diffraction image in Figure 2(c) corresponds to a strongly deformed, capsule-shaped droplet with $AR = 1.95$. The capsule shape is indicated by the small curvature of the streak, as discussed earlier [21, 23].

All images in Figure 2 exhibit blank horizontal stripes along their middle sections. These result from the gap between the upper and the lower panels of the pnCCD detector. Vertical stripes on the lower panel are caused by imperfect data readout for strong diffraction images.

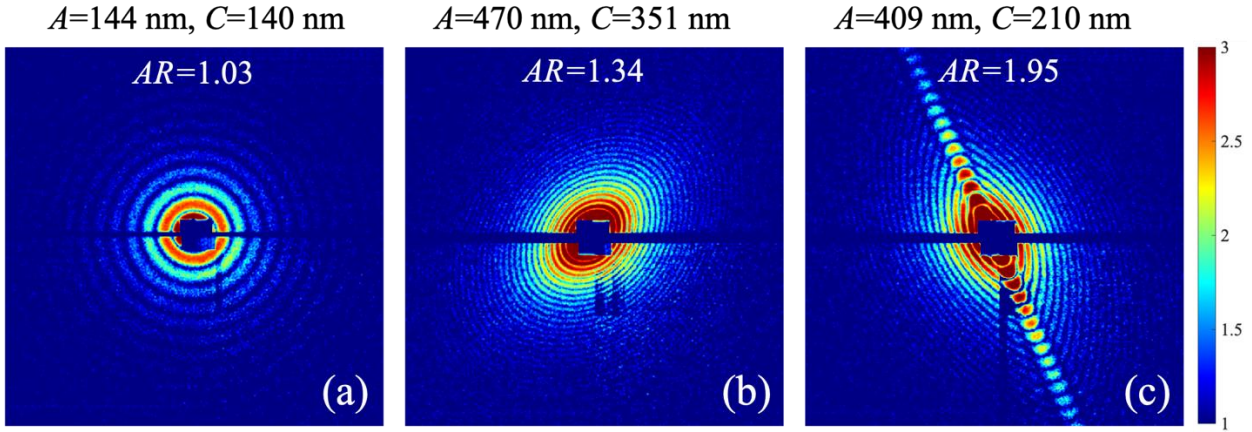

**Figure 2.** Diffraction patterns of pure $^3$He droplets shown on a logarithmic color scale as indicated on the right. Images represent the central 660 × 660 detector pixels. Corresponding droplet projection half-axes ($A$, $C$) and their aspect ratio ($AR$) are displayed at the top of each image.

During the measurements, approximately 900 intense diffraction images from pure $^3$He droplets are obtained, each providing a unique set of $A$ and $C$ values. Similar measurements are performed for $^4$He droplets, providing ~300 patterns as an independent reference for comparison. The measurements for a given isotopic fluid do not exhibit any systematic variation with temperature, thus, the results obtained at different temperatures are combined to improve statistics. Figure 3(a) displays the measured distribution of the droplet's long axis, $A$, for $^3$He and $^4$He droplets, as represented by blue and red bars, respectively. The average sizes of $^4$He droplets are approximately a factor of two larger than those of $^3$He droplets. The sizes of the $^3$He droplets vary between $A = 52$ nm and $A = 796$ nm with an average of $<A> = 162$ (94) nm, whereas $^4$He droplets exhibit a larger spread, ranging from $A = 55$ nm to $A = 1250$ nm with an average of $<A> = 355$ (260) nm. Throughout this article, values in parenthesis give the root mean square deviation of the corresponding quantity. Figure 3(b) shows the $AR$ distribution for $^3$He and $^4$He droplets. The largest $AR$s are 1.99 for $^3$He and 1.72 for $^4$He. The $AR$ histograms for the two isotopes appear more similar than the size distributions, with the vast majority of droplets exhibiting $AR < 1.2$. The average $AR$s for $^3$He and $^4$He droplets are very close with $<AR>_3 = 1.055$ (0.082) and $<AR>_4 = 1.067$ (0.089), respectively. We also found that at $AR < 1.4$, where more than 98% of the droplets were found, the distribution of the *AR*-1 values is well approximated by an exponential. The number of detected droplets with *AR* > 1.4 is too few to determine the distribution. Figure 3(c) shows the *AR* vs. half axis *A* for all data points used to produce Panels (a) and (b). The results for $^3$He and $^4$He are shown by blue stars and red circles, respectively. It is readily apparent that for both isotopes, the fraction of droplets with large *AR* (> 1.2) is higher in larger droplets.

In contrast to the temperature-independent droplet sizes reported here, previous measurements on $^4$He droplets found continuously increasing sizes with decreasing temperature [16]. At $T_0 <$ 4 K and $P_0$ = 20 bar, $^4$He expansion leads to the formation of a jet that breaks up into micron-sized droplets due to Rayleigh instability. [16, 49] This mechanism gives rise to an extremely collimated beam of droplets, the occurrence of which was not observed during this work with either $^3$He or $^4$He. We conclude that, most likely, the flow through the nozzle was affected by imperfections such as microscopic damage or partial obstruction of the nozzle by impurities. Previous experiments with $^4$He droplets in our group demonstrated that, under such conditions, decreasing the nozzle temperature below a certain value does not result in any increase in average droplet size [50], which is in agreement with the observations in this work.

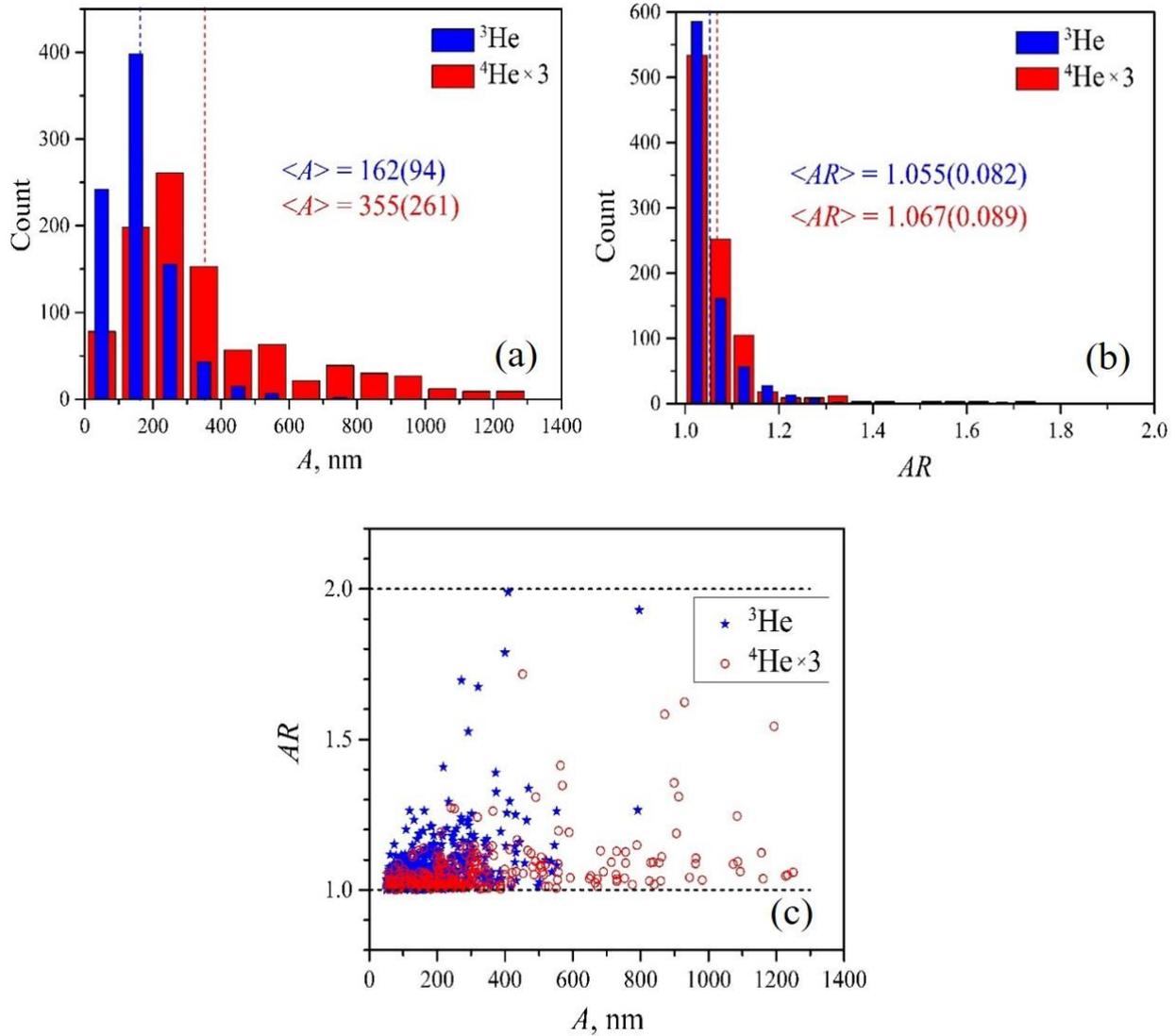

**Figure 3.** Droplet size (a) and aspect ratio (b) distributions for $^3$He (blue) and $^4$He (red) isotopes. Corresponding average values as obtained from the entire data sets (with root mean square deviations in parentheses) are listed in each plot and indicated by vertical dashed lines. The results for $^4$He were multiplied by a factor of 3 for ease of comparison, as the total number of diffraction images obtained for $^3$He and $^4$He were ~900 and ~300, respectively. Panel (c) shows the *AR* vs. half axis *A* for all data points used to produce (a) and (b). The results for $^3$He and $^4$He droplets are shown by blue stars and red circles, respectively.

## 4. Discussion

### 4.1. Droplet size distribution

Figure 3(a) shows that the observed droplet size distributions peak at some small values of *A*, decrease sharply towards smaller *A*, and decrease more gradually towards larger *A*. The measured distribution reflects the actual distribution in the beam multiplied by the probability to detect a droplet of value *A* in the diffraction experiment. In the SM [45], it is shown that for a spherical droplet, the detection probability scales as $ln\frac{R^4}{R_0^4}$ if $R \geq R_0$ and 0 if $R < R_0$, where R is the radius of the droplet and $R_0 \approx 50$ nm is the radius of the smallest detectable droplet. Accordingly, for $R \gg R_0$ the detection probability is a slowly changing logarithmic function of *R* and can often be ignored. However, as *R* approaches $R_0$, detection probability goes to zero, which explains the decrease of the counts at small *A* in Figure 3(a).

In the literature, droplet size distributions are usually discussed in terms of the number of atoms per droplet, owing to the detection technique, which is often based on mass spectroscopy [51]. Figure 4 shows the size distribution for $^3$He droplets in a logarithmic representation. For $N_3 < 2\times10^9$, the $^3$He droplet size distribution is approximately exponential, $P(N_3) = \frac{S \cdot \Delta}{<N_3>} \exp(-\frac{N_3}{<N_3>})$, with $<N_3> \approx 5.6 \times 10^8$ and *S* being the total number of the detected droplets and $\Delta$ is the bin size. For $N_3 > 2\times10^9$, the probability of detecting droplets becomes greater than predicted by an exponential dependence and extends to very large droplet sizes, up to about $N_3 = 2\times10^{10}$ (not shown). An exponentially declining size distribution was also found in a recent study of $^4$He droplets at the FERMI FEL.[23] On the other hand, the size distribution of smaller $^3$He droplets with $N_3 \leq 10^7$ obtained at $P_0 = 20$ bar and $T_0 \geq 5$ K was found to be close to log-normal [44].

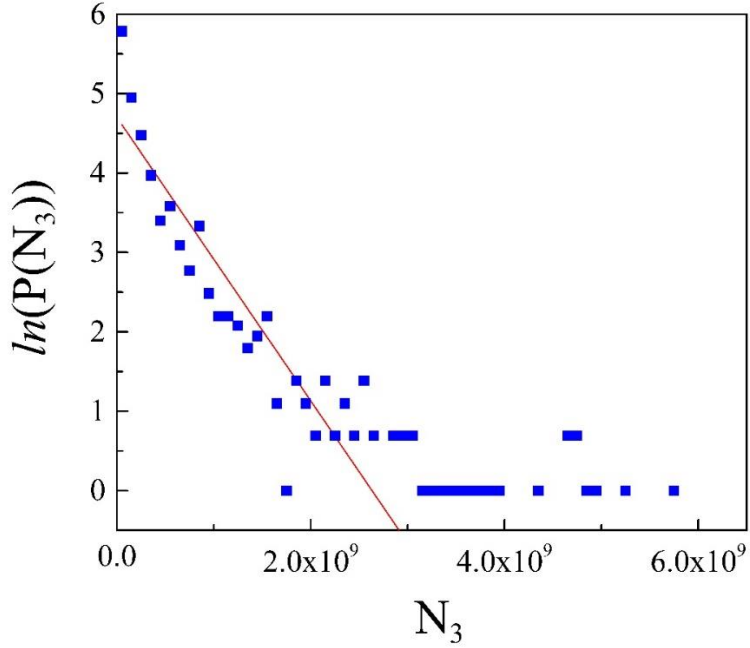

**Figure 4.** Size distribution of $^3$He droplets. The red line represents an exponential distribution. See text for details.

**4.2. Droplet aspect ratio**

The aspect ratios *AR* provide access to the angular momentum and angular velocity of the droplets. Unfortunately, the actual aspect ratio, $ar = a/c$, cannot be obtained from the individual diffraction images at small scattering angle due to the unknown orientation of the droplets with respect to the X-ray beam. However, one can obtain the average actual aspect ratio $\langle ar \rangle$ from the average apparent aspect ratio $\langle AR \rangle$ assuming a random droplet orientation as described in the following.

In classical droplets, the largest aspect ratio of stable, axially symmetric droplets is $ar = 1.47$ [52, 53]. About 99% of the measurements in Figure 3 have $AR < 1.4$ in agreement with previous measurements in $^4$He droplets [18, 21-23]. Here, we assume that the overwhelming majority of droplets with $AR < 1.4$ have oblate, axially symmetric shapes. We also assume that the data contain less than ~10 events from prolate $^3$He droplets that are oriented in such a way that their projections yield $AR < 1.4$ but, within the accuracy of the data analysis, the corresponding diffraction images cannot be distinguished from those for oblate droplets. This estimate is based on the number of events producing $AR > 1.4$, which are entirely ascribed to prolate droplets. For shapes with $AR < 1.4$, the average values for the observed major half axis $A$ and aspect ratio $AR$ of: $<A_3> = 160 \pm 3$ nm, $<AR_3> = 1.049 \pm 0.003$, $<A_4> = 348 \pm 14$ nm, $<AR_4> = 1.059 \pm 0.005$.

To translate the measured $<AR>$ into the actual $<ar>$, we assume a spheroid with a well-defined $ar$ and calculate its projection on the detector plane when its symmetry axis $c$ subtends an angle $\alpha$ with the normal to the plane. The aspect ratios of the diffraction pattern ($AR$) and of the spheroid of which the x-rays diffract ($ar$) are related by: $AR = \sqrt{cos^2(\alpha) + ar^2 sin^2(\alpha)}$, see eq. (S2.8) in the SM to [18]. The average $AR$ of an ensemble of randomly aligned droplets is then calculated as $\langle AR \rangle = \int_0^{\frac{\pi}{2}} AR(\alpha) \cdot sin(\alpha) \cdot d\alpha$, where $sin(\alpha)$ represents the probability of detecting a spheroid at angle $\alpha$. Calculations have been performed for spheroids with a variety of $ar$. They show that, in the range $1 \leq ar \leq 1.4$, $\langle AR \rangle$ scales nearly linearly with $ar$ according to $\langle AR \rangle - 1 = 0.64(ar - 1)$. Due to the linear relationship between $<AR>$ and $ar$, the same formula also applies when considering not just an orientation-averaged ensemble with one specific $ar$, but also averages over all orientations and all true aspect ratios $ar$: $\langle AR \rangle - 1 = 0.64(\langle ar \rangle - 1)$. From this relationship, the average true aspect ratios for $^3$He and $^4$He droplets are derived as $\langle ar \rangle_3 = 1.077$ (0.005) and $\langle ar \rangle_4 = 1.092$ (0.08), respectively.

Since the projection $A$ always assumes the value $a$ for an axisymmetric droplet, we can determine the true average major half axes $\langle a \rangle$ from the measurements as $\langle a_3 \rangle = 160$ (90) nm and $\langle a_4 \rangle = 348$ (254) nm for $^3$He and $^4$He droplets, respectively. Using the approximation $\langle ar \rangle \approx \langle a \rangle / \langle c \rangle$ and the results summarized in Figure 3 with $AR$ less than 1.4, the average minor half axis $\langle c \rangle$ for $^3$He and $^4$He is determined as $\langle c_3 \rangle = 150$ (80) nm and $\langle c_4 \rangle = 300$ (200) nm, respectively.

With the obtained $\langle ar \rangle$, $\langle a \rangle$, and $\langle c \rangle$, the average number of He atoms in the droplet $\langle N_{3,4} \rangle$ can be deduced as, $\langle N_{3,4} \rangle = \langle V \rangle \times n_{3,4}$, where $\langle V \rangle = \frac{4 \cdot \pi \cdot <a>^2 \cdot <c>}{3}$ is the volume of an oblate spheroid, and $n_{3,4}$ is the number density of liquid $^3$He and $^4$He at low temperature, with values of $n_3 = 1.62 \times 10^{28}$ m$^{-3}$ [38] and $n_4 = 2.18 \times 10^{28}$ m$^{-3}$ [41, 42], respectively. The average sizes for droplets using the above calculated $\langle a \rangle$ and $\langle c \rangle$ values with aspect ratios less than 1.4 are $<N_3> = 2.6 \times 10^8$ and $<N_4> = 3.5 \times 10^9$.

## 4.3. Average angular momenta and angular velocities of $^3$He and $^4$He droplets

As previously described for $^4$He droplets [18, 21-23], we assign the shape deformation in $^3$He droplets to centrifugal distortion. It has been reported that the shapes of rotating $^4$He droplets closely follow the equilibrium shapes of classical droplets having the same values of angular momentum [21, 23, 26, 27]. This is also expected to be the case for $^3$He droplets, which at the temperature of these experiments (~ 0.15 K) [35, 54], should behave classically because of the high viscosity of about 200 µP and small mean free path (a few nm) of elementary excitations at this temperature [38]. In the recent density functional calculations the shapes of rotating $^3$He droplets were found to be very close to those predicted for classical droplets [39]. The blue curve in Figure 5 shows the stability diagram of the classical droplets in terms of the reduced angular momentum (Λ) and the reduced angular velocity (Ω), which are given by [52, 53],

$$\Lambda = \frac{L}{\sqrt{8 \cdot \sigma \cdot \rho \cdot R^7}} \qquad (1)$$

$$\Omega = \sqrt{\frac{\rho \cdot R^3}{8 \cdot \sigma}} \cdot \omega \qquad (2).$$

Here, L and ω are the angular momentum and angular velocity, respectively, σ is the surface tension of the liquid, ρ is the liquid mass density, and R is the droplet radius in a quiescent state. For liquid $^4$He and $^3$He at low temperature, the surface tensions are $\sigma_4$ = 3.54·10$^{-4}$ N/m [42] and $\sigma_3$ = 1.55·10$^{-4}$ N/m [55], respectively, corresponding to densities of $\rho_4$ = 145 kg/m$^3$ [42] and $\rho_3$ = 82 kg/m$^3$, [56]. With increasing Λ, the droplet's equilibrium shape transitions from spherical to oblate axially symmetric. At Ω ≈ 0.56, Λ ≈ 1.2, *ar* ≈ 1.47, the stability curve bifurcates into two branches; an unstable upper branch (dashed blue curve) representing axially symmetric droplets and a stable lower branch (dotted blue curve) representing prolate triaxial droplets. The stable prolate branch represents triaxial ellipsoidal and capsule shaped droplets with 1.2 < Λ < 1.6, and dumb-bell shaped droplets at Λ > 1.6 [21, 23, 52, 53]. For Λ > 2, droplets become unstable and break up. Also shown in Figure 5 is the aspect ratio (*a*/c) of droplets along the axisymmetric branch as a function of Λ, which is represented by the red curve [21]. Using an exponential distribution of the AR values: $P(AR - 1) = \frac{1}{<AR-1>} \exp\left(-\frac{AR-1}{<AR-1>}\right)$ and the curves in Figure 5, the average Λ for

$^3$He and $^4$He is obtained as 0.40 and 0.44, indicated as green circles and black crosses, respectively. From the stability diagram, we obtain $\Omega$ for $^3$He and $^4$He as $\Omega_3 = 0.24$ and $\Omega_4 = 0.26$, respectively. Using Equations (1, 2), the angular momentum ($L$) is obtained as $L_3 = 1.3\times10^9$ $\hbar$ and $L_4 = 5.3\times10^{10}$ $\hbar$ for the average sized $^3$He and $^4$He droplets, respectively. Similarly, $L$ per atom of the droplet is obtained as 4.9 $\hbar$ and 16.3 $\hbar$ for $^3$He and $^4$He droplets, respectively, and ω was calculated as $1.8\times10^7$ rad/s and $6.1\times10^6$ rad/s for $^3$He and $^4$He, respectively. Although the $^4$He droplets and $^3$He droplets have similar <Λ>, $^4$He droplets have about a factor of three larger $L$ per atom. Mathematically, this effect stems from the different of the factors of $\sqrt{\sigma \cdot \rho \cdot R^7}$ in Equation (1) in $^3$He and $^4$He droplets.

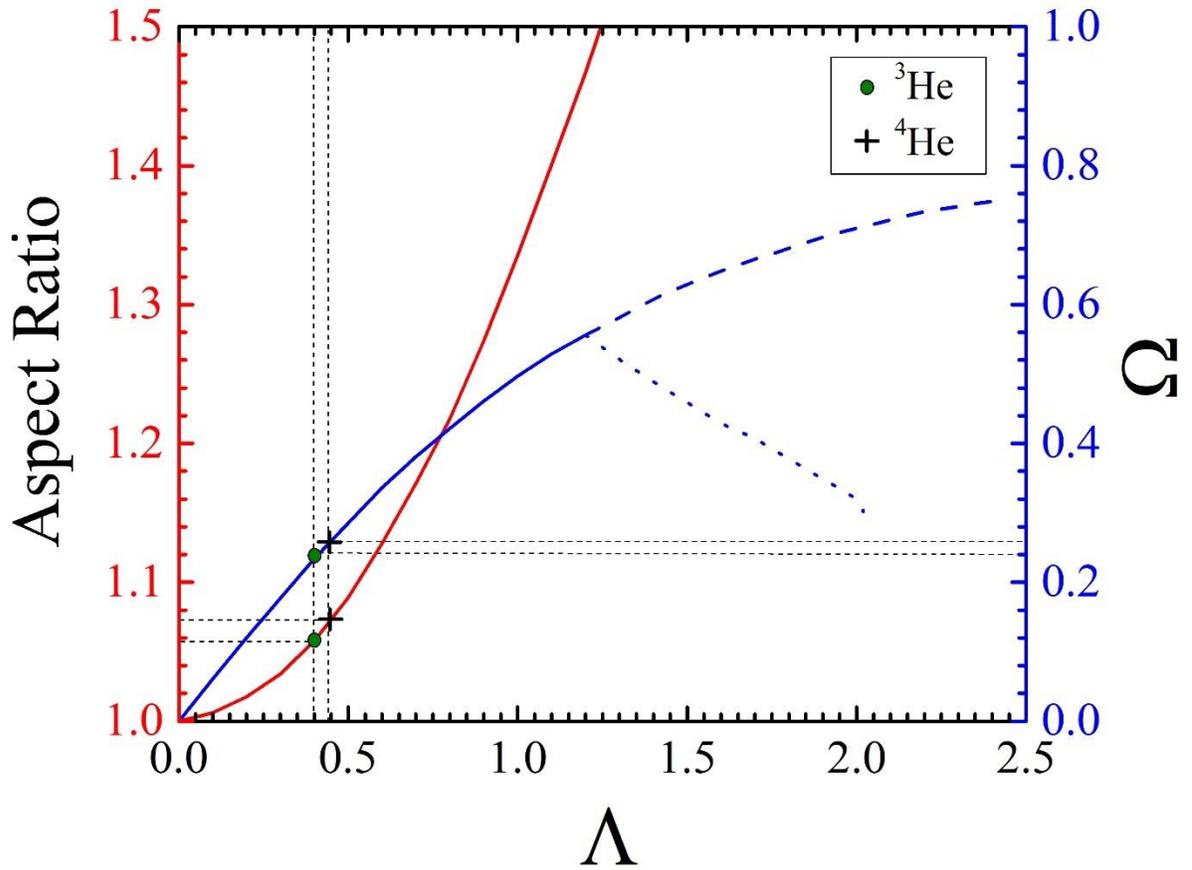

**Figure 5.** Red curve: Calculated aspect ratio as a function of reduced angular momentum (Λ) for axially symmetric oblate droplet shapes. Blue curve: stability diagram of rotating droplets in terms of reduced angular velocity (Ω) and reduced angular momentum (Λ). The upper branch (dashed blue) corresponds to unstable axially symmetric shapes. The lower branch (dotted blue) is associated with prolate triaxial droplet shapes resembling capsules and dumbbells. The green circle and black cross on the red curve represent the average ⟨ar⟩ for ³He and ⁴He droplets, respectively, obtained in this work (with AR < 1.4). Similar markers on the blue curve indicate the (Ω, Λ) values for the average droplets.

### 4.4. Formation of rotating droplets in the fluid jet expansion

It is remarkable that in spite of their very different physical properties, $^3$He and $^4$He droplets have, on average, very similar values of $\Omega$ and $\Lambda$. Previous XFEL experiments with $^4$He droplets yielded average aspect ratios, <AR>, in the range of 1.06 – 1.08 at $P_0$ = 20 bar and $T_0$ = 4 - 7 K, which spans average droplet sizes from 200 nm to 1000 nm in diameter (see Figure 4.11 in Reference [50]). Thus, it is noteworthy that very similar average aspect ratios, and therefore $\Omega$ and $\Lambda$, were obtained at different $T_0$. Comparable <AR> were obtained in experiments involving different nozzle plates, including measurements with partially obstructed and intact nozzles [50]. Hence, it seems that the acquired <AR> is largely independent of particular nozzles used in the experiments. Similar results for non-superfluid $^3$He and superfluid $^4$He droplets indicate that the state of the droplets has a small effect on the resulting average angular momentum.

In previous works [21, 57], we have conjectured that during the passage of fluid helium through the nozzle, the fluid interacts with the nozzle channel walls and acquires vorticity, which is eventually transferred to the droplets. Accordingly, the droplets' angular velocity may be estimated based on the nozzle diameter ($d$) of 5 μm and the measured $^4$He droplet beam velocity of $v$ = 170 m/s [16]. Using Bernoulli's equation, the velocity of $^3$He droplets can be estimated to be $v \approx$ 225 m/s. If the fluid at the center of the nozzle moves with the beam velocity and falls linearly to zero at the walls, the resulting velocity gradient gives an estimate for the average vorticity of the fluid as 2·$v/d$. The average vorticity will be up to about a factor of two larger for a more realistic velocity profile with a sub-linear change of velocities close to the nozzle center. One can also assume that, upon breakup of the fluid into droplets, vorticity is conserved and, thus, the angular velocity of the droplets can be obtained as half of the average vorticity: $\omega$ = $v/d$. Accordingly, the estimated average angular velocity of $^4$He and $^3$He droplets is $3.4 \times 10^7$ rad/s and $4.5 \times 10^7$ rad/s, respectively. Such high angular velocities can only be sustained by rather small droplets.

It is challenging to explain the similarities in reduced angular velocity and angular momentum in $^3$He and $^4$He droplets based on the stability diagram in Figure 5 and the estimated vorticities. Moreover, the size and shape distributions in Figure 3 as observed at high vacuum far downstream (~1 m) from the nozzle originate from processes in the high-density region inside or close to the nozzle, where collisions between droplets with the dense He gas must play an important role. For example, for a droplet 300 nm in radius, rotating at $10^7$ rad/s, the peripheral velocity will be ~3 m/s. In the regime of extensive jet atomization as in this work, a large spread of droplet velocities up to $\Delta v/v$ ~5% has previously been observed [58]. Thus, with a characteristic droplet velocity on the order of 200 m/s, the droplets may have significant relative collision velocities of ~10 m/s, which are sufficient to produce rapidly spinning products. Further downstream, presumably a few mm away from the nozzle, the number density of the gas and droplets decrease, the collision rates decrease, and the angular momenta of individual droplets remain constant further downstream.

Although we are currently unable to provide a quantitative model of the processes close to the nozzle, it is instructive to consider the evolution of a droplet driven at some angular velocity as opposed to free droplets with a constant angular momentum. The corresponding driving force may originate from the aforementioned collisions. The prolate branch on the stability curve of driven droplets is unstable at constant ω [53]. Driven droplets will climb along the axially symmetric branch until they reach the bifurcation point at $\Omega = 0.56$ (Figure 5) at which point they will enter the unstable prolate branch. Here, further elongation of the droplets occurs, culminating in the fission and formation of two nearly spherical droplets, each having one half the volume of the parent droplet [53]. The entire angular momentum at the fission point is due to the relative motion of the daughter droplets, as there is no rotation within either of them. Similar to the parent droplets, the daughter droplets will acquire angular momentum via collisions. The fission cycle continues until sufficiently small, stable droplets are formed or the droplets are far away from the nozzle, where the driving force diminishes. Because the occurrence of such a cycle is largely independent of the choice of He isotope, the process should yield very similar values of <AR>, <Λ> and <Ω>, independent of the droplet size and composition. This model could explain why similar Ω and Λ were obtained at different $T_0$.

From the average nozzle vorticity and using $\Omega = 0.56$ from the bifurcation point (from Equation (2)), the radius of the largest stable droplets can be estimated to $R_3 = 130$ nm and $R_4 = 170$ nm for $^3$He and $^4$He droplets, respectively. The obtained critical radii are larger than the smallest observed droplets of about 50 nm in radius. This may indicate that the fission process yields smaller droplets than our idealized estimate. For the above estimates, we have applied low-temperature surface tension values, which are lower at higher temperatures inside the nozzle, leading to smaller radii. At the nozzle, in addition to rotation, the droplets will likely experience shape oscillations that may decrease the threshold Ω for turning into prolate shapes, eventually leading to fission. In addition, droplet sizes also decrease due to evaporation in vacuum.

## 5. Conclusions

In this work, bosonic $^4$He and, for the first time, fermionic $^3$He droplets are studied by single-pulse X-ray coherent diffractive imaging. Statistics of the droplets' sizes, aspect ratios, reduced angular momenta and reduced angular velocities are compared for superfluid $^4$He droplets and normal fluid $^3$He droplets. Since the experiments only give access to projections of droplets onto the detector plane, estimates are made to determine the true average axes and aspect ratios. It is found that, although the superfluid droplets have a much higher average angular momentum, the two kinds of droplets have very similar average aspect ratios and, thus, similar average reduced angular momenta and reduced angular velocities. This surprising result may result from the formation of the droplets through turbulent nozzle flow and the atomization regime in immediate vicinity of the nozzle. We conjecture that the droplets' rotation is driven by a combination of the liquid flow velocity gradient inside the nozzle and collisions close to it, leading to elongation and, ultimately, fragmentation into daughter droplets, which may undergo repeated collision-elongation-fragmentation cycles.

Future studies will shed more light on the origin of angular momentum in droplets produced via fragmentation of a fluid. A large number of studies discusses the fragmentation of classical liquids upon jet expansion [59, 60]. However, to the best of our knowledge, the amount of angular momentum contained in the resulting droplets remained unknown. It is therefore interesting that the jet atomization of classical liquids produces highly rotating droplets similar to quantum He droplets.

The availability of the large $^3$He droplets suitable for single-pulse diffraction experiments also opens additional research directions. Vortex-induced cluster aggregation has so far been unique to superfluid $^4$He. It is of high interest to expand experiments to non-superfluid $^3$He and study the aggregation patterns in rotating fermionic droplets. Dopant aggregation mechanisms and the morphology of the phase separation in rotating mixed $^3$He/$^4$He droplets presents another frontier. [61]

## 6. Acknowledgements

This material is based upon work supported by the National Science Foundation under Grant No. DMR-1701077 (A.F.V.). C.S., B.W.T., M.B. and O.G. are supported by the U.S. Department of Energy, Office of Science, Office of Basic Energy Sciences, Chemical Sciences, Geosciences and Biosciences Division, through Contract No. DE-AC02-05CH11231. M.B. acknowledges support by the Alexander von Humboldt foundation. Portions of this research were carried out at the LCLS, a national user facility operated by Stanford University on behalf of the U.S. DOE, OBES under beam-time Grant No. LU46: Molecular Self-Assembly Close to 0 Kelvin. Use of the Linac Coherent Light Source (LCLS), SLAC National Accelerator Laboratory, is supported by the U.S. Department of Energy, Office of Science, Office of Basic Energy Sciences under Contract No. DE-AC02-76SF00515. We thank Manuel Barranco for carefully reading the manuscript and making Ref. [39] available to us prior to publication.

**Supplementary material to:**

**Shapes of rotating normal fluid $^3$He versus superfluid $^4$He droplets in molecular beams**

Deepak Verma[1], Sean M. O. O'Connell[1], Alexandra J. Feinberg[1], Swetha Erukala[1], Rico M. Tanyag[1,2], Charles Bernando[3,4], Weiwu Pang[4], Catherine A. Saladrigas[6,7], Benjamin W. Toulson[6], Mario Borgwardt[6], Niranjan Shivaram[8,9], Ming-Fu Lin[8], Andre Al Haddad[10], Wolfgang Jäger[11], Christoph Bostedt[10,12], Peter Walter[8], Oliver Gessner[6*] and Andrey F. Vilesov[1,3*]

[1] Department of Chemistry, University of Southern California, Los Angeles, California 90089, USA
[2] Max-Born-Institut für Nichtlineare Optik und Kurszeitspektroskopie, Max-Born-Straße 12489, Berlin, Germany
[3] Department of Physics and Astronomy, University of Southern California, Los Angeles, California 90089, USA
[4] School of Information Systems, BINUS University, Jl. K.H. Syahdan No. 9, Kemanggisan, Palmerah, Jakarta 11480 Indonesia
[5] Viterbi School of Engineering, University of Southern California, Los Angeles, California 90089, USA
[6] Chemical Sciences Division, Lawrence Berkeley National Laboratory, Berkeley, California 94720, USA
[7] Department of Chemistry, University of California, Berkeley, California 94720, USA
[8] LCLS, SLAC National Accelerator Laboratory, 2575 Sand Hill Road, Menlo Park, California 94025, USA
[9] Department of Physics and Astronomy, Purdue University, West Lafayette, Indiana 47907, USA
[10] Laboratory for Femtochemistry (LSF), Paul Scherrer Institut, 5232 Villigen-PSI, Switzerland
[11] Department of Chemistry, University of Alberta, Edmonton, Alberta T6G 2G2, Canada
[12] LUXS Laboratory for Ultrafast X-ray Sciences, Institute of Chemical Sciences and Engineering, École Polytechnique Fédérale de Lausanne (EPFL), CH-1015, Lausanne, Switzerland

Present Address:

[⊥] OVO (PT. Visionet Internasional), Lippo Kuningan 20th floor, Jl. HR Rasuna Said Kav. B-12, Setiabudi, Jakarta, 12940, Indonesia
[#] Department of Physics and Astronomy, Purdue University, West Lafayetta, Indiana 47907, USA

March 9th, 2020

## S1. Detection probability of He droplets vs droplet size

Fig. 3(a) of the main text shows that the observed droplet size distributions peak at some small values of $A$. The measured distribution reflects the actual distribution in the beam multiplied by the probability, $DP(R)$ that a droplet with a major half axis $A$ will be detected in the diffraction experiment. To estimate the actual size distribution, we deduce a simple, idealized detection probability function for spherical droplets of radius $R$. As most of the detected droplets have aspect ratios close to 1, the formula for a sphere provides a good first-order estimate for the detection probability distribution. According to the Rayleigh - Gans approximation, [1, 2] the total scattering intensity from a spherical droplet of radius $R$ is given by:

$$I(R) = 8\pi \cdot R^4 \left|\frac{n^2-1}{\lambda}\right|^2 \cdot \Phi = C \cdot R^4 \cdot \Phi. \tag{1}$$

Here, $n$ is the complex refractive index of liquid $^3$He or $^4$He, $\lambda$ is the wavelength of the scattered light, and $\Phi$ is the photon flux of the incoming X-ray beam. The total measured scattering intensity needs to be higher than a certain threshold value, $I_{th}$, for the droplet to be detected. We disregard any intensity loss due to the central cut in the detector panels and the gap between them and assume that the X-ray beam has a Gaussian intensity profile:

$$\Phi(r) = \Phi_0 \exp\left(-\frac{2r^2}{w_0^2}\right) \tag{2}$$

Here, r is the distance from the beam axis perpendicular to the direction of the X-ray beam and $w_0$ is the beam waist. Here we assumed that R<<$w_0$. The value of $I_{th}$ determines the smallest droplet size, $R_0$, that may be detected for a droplet residing on the beam axis ($r = 0$):

$$I_{th} = C \cdot R_0^4 \cdot \Phi_0 \tag{3}$$

Larger droplets with radius $R > R_0$ may also be detected if they reside off axis at radii smaller than $r_{max}$ with

$$r_{max} = \left(\frac{w_0^2}{2} \ln \frac{C \cdot R^4 \cdot \Phi_0}{I_{th}}\right)^{1/2} = \left(\frac{w_0^2}{2} \ln \frac{R^4}{R_0^4}\right)^{1/2} \tag{4}$$

Accordingly, the detection probability $DP(R)$ is proportional to the area within $r_{max}$, which grows $\propto \ln \frac{R^4}{R_0^4}$. In this work $R_0 \approx 50$ nm. For $R >> R_0$ the detection probability is a slowly changing logarithmic function of $R$ and can often be ignored. However, as $R$ approaches $R_0$, $DP(R)$ goes to zero, which explains the decrease of the counts at small $A$ in Fig. 3(a) of the main text.

## S2. $^3$He Recycling System

Due to the considerable cost of $^3$He, a recycling system is employed during the experiments. The design of the gas recycling system was inspired by a similar system used for experiments with $^3$He droplets [3-5] and for $^3$He gas circulation systems in dilution refrigerators [6]. The $^3$He recycling system fulfills the following functions: 1) collection of recycled gas, 2) cleaning of recycled gas, 3) pressurization of clean gas, and 4) storage of clean gas.

Figure S1 shows a schematic of the recycling system. Two, inward-facing triangles denote valves. Blue arrows in Fig. S1 indicate the direction of the helium flow during operation. The gas exits the cryogenic nozzle, is pumped by turbo pumps that are backed by scroll pumps (Leybold SC 30D and Anest Iawata ISP 250C). An Adixen DFT-25 microfiber-based dust filter is installed at the exit of the scroll pump to stop debris from entering the system. Gas is collected from the output of the scroll pump and impurity gases are frozen out on LN$_2$-cooled zeolite traps. The purified helium gas is compressed by a Fluitron S1-20/150 compressor and resupplied to the cryogenic nozzle. When not in operation, gas can be stored in the cylinders shown in the upper right portion of Figure S1.

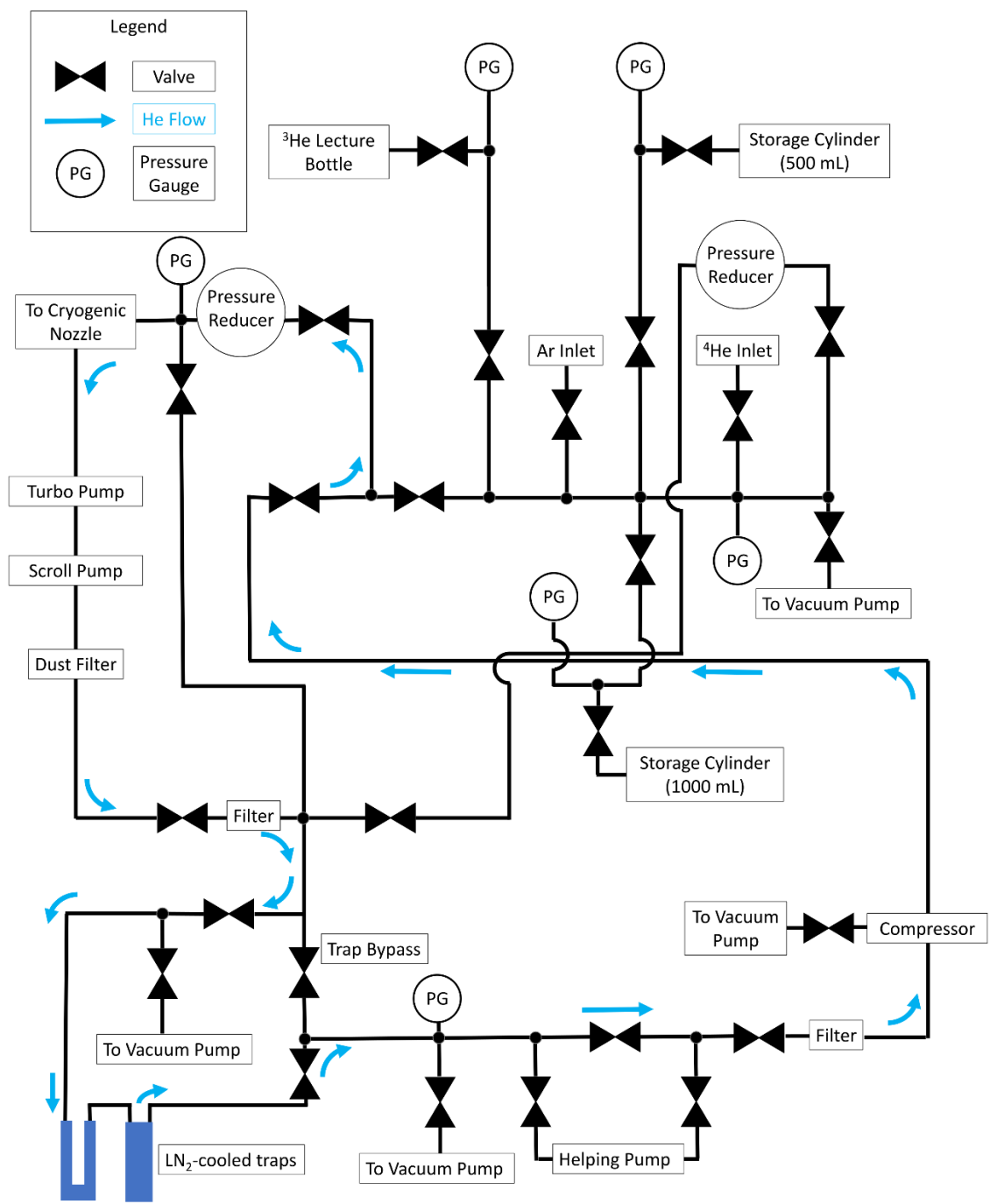

**Fig. S1**. A schematic of the $^3$He recycling system. The black lines indicate tubing connections between valves and other parts of the system. Blue arrows indicate the direction of helium flow during operation. The system can be evacuated before operation to preserve the purity of the gas. Connections to external vacuum pumps are indicated by labels "To Vacuum Pump."